\documentclass[review,sort&compress]{elsarticle}

\usepackage{lineno,hyperref,amsmath,amssymb}
\modulolinenumbers[5]

\begin{document}

\begin{frontmatter}

\title{Non-Convergent Perturbation Theory and Misleading Inferences about Parameter
Relationships: the Case of Superexchange}

\author{Demetra Psiachos}
\address{Crete Center for Quantum Complexity and Nanotechnology, Department of Physics, University of Crete, Heraklion 71003, Greece}
\ead{dpsiachos@gmail.com}

\begin{abstract}
We discuss the well-known three-center cation-anion-cation model for superexchange in
insulating transition-metal compounds using
limiting expansions for the Anderson-Hubbard model. We find that due to the three
interfering energy scales in the model, a limiting expression for the superexchange $J$ for the
idealized Mott-Hubbard (M-H) case $t\ll U\ll \Delta$ cannot be formally defined. We further show that no
single expansion variable can describe any type of limiting behaviour for superexchange. The well-known
$t^4$ expression for M-H insulators, obtained from path-dependent series expansions, is not unique to these
systems as it can also be obtained with many other different expansions, in which
either the $d-p$ energy difference $\Delta$ or the $d$-electron correlation $U$ can
actually be small. At times, and particularly for milder relationships between the parameters,
$t\lesssim U\lesssim\Delta,$ the reverse form of the series expansions can yield
better agreement with the exact results. This implies that the fitting of experimental
data to the simple expressions derived from path-dependent series expansions
can lead to qualitatively incorrect relationships between the parameters,
fictitiously within the M-H regime.
\end{abstract}


\end{frontmatter}

\linenumbers

\section{Introduction}

The problem of superexchange, or the exchange interaction of
two separated electron spins, is an old one, dating back \textit{e.g.} to the work of Anderson~\cite{Anderson}
who obtained a simple expression for describing the interaction of $3d$ spins in Mn cations separated
by a filled shell of O $2p$ electrons. Since then the concept has been applied to the estimation of
parameters in a wide variety of systems, such
as hopping parameters in transition-metal compounds,~\cite{Jefferson1,Jefferson2} high T$_c$
superconductors,~\cite{Manousakis} and electron transfer rates in donor-acceptor systems, such as radical pairs in biological molecules~\cite{rps}.
Following Anderson, many works in the literature have presented slight modifications and extensions
of the superexchange expression $J$~\cite{Geertsma,Zaanen1,Zaanen2,prb2007}. Experimentally,
the electronic parameters cannot all be measured directly. Some studies,~\cite{SpinWave,feldkemper} which rely
on assumed models in order to fit experimental data on exchange,
typically obtained from spin-wave~\cite{SpinWave1,SpinWave2} or magnetic-susceptibility measurements~\cite{magsusc} (for a review
of experiments see Ref.~\cite{deJonghAdvPhys}), are biased by the suitability of the model. In contrast, some other analyses,
which fit photoemission spectra
to parameters from cluster models~\cite{LeeOh,Bocquet92,Bocquet}, constitute an independent approach to obtaining values
for the electronic parameters.

Indeed it has been shown, based on cluster model calculations including a full description
of configuration interactions, that the late $3d$ transition-metal oxides should not be
characterized as Mott-Hubbard (M-H) insulators, in which the energy gap is determined by the $d-d$ electron
correlation $U$ on the metallic species. Instead, these systems have been reclassified
as charge-transfer (C-T) insulators~\cite{LeeOh,Bocquet92} in which
the gap is determined by the energy of charge-transfer $\Delta$ between metal and insulator species.
Similarly, as a result of comprehensive cluster calculations, the early
transition-metal oxides are now considered to be at least intermediate
between the two regimes, with some classified as C-T~\cite{Bocquet}.

In this paper we rederive the well-known problem of the superexchange interaction of singly-occupied cations
in a three-site cluster model as a proof-of-principle that expressions for limiting forms are
generally path-dependent
whenever multiple energy scales are present. This concept is generalizable to the plethora of
other problems in physics desribed by multiple energy scales and
for which analytical expressions are only available in limiting forms.

Obtaining a simple expression for complex quantities often relies on series expansions. A
careful examination of the convergence properties of the expansion is necessary whenever there
are multiple scales in the problem. Bender~\cite{Bender} has recently revisited
the problem of correlated limits of multiple energy scales, in which the same variable is involved
in more than one limit, in terms of $\mathcal{PT}$ symmetry, while
a recent study studying correlated limits
in $\mathcal{PT}$-symmetric systems~\cite{Psiachos2014}
numerically found that the apparent phase depends on the
path taken towards the limits.

We show that the two expressions comprising the expression for superexchange, the singlet
and triplet energies, possess different convergence properties, and as such,
no single variable can reliably describe the limiting behaviour for superexchange. We also find that the series
expansions used to obtain the limiting behaviour for superexchange are non-commuting, and that in a large
part of parameter space, the reverse from the usual sequence of expansions yields better agreement with the exact expression.
In Sec.~\ref{sec:methods} we outline our method of
exact solution, and compare it with other methods, which are based on perturbation theory. The exact solution is
expanded in different ways in Sec.~\ref{sec:results} and compared with well-known limiting forms from the literature. A
discussion is provided to explain the root of the discrepancies, and finally a summary is given
in Sec.~\ref{sec:conclusions}.

\section{Theoretical Methods}
\label{sec:methods}
\subsection{Model of the system}
\begin{figure}[htb]
\includegraphics[width=3cm]{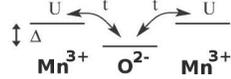}
\caption{Site-centred Wannier orbitals in MnO and their associated energies. The site energy difference $\Delta$ (charge-transfer
energy) is defined
as $\epsilon_d-\epsilon_p$. For double occupation, there is a U only for the Mn $d$ orbitals and none
for the O $p$ orbital.}
\label{superlevels}
\end{figure}

We specifically consider the Wannier-orbital representation of
Anderson's model of the MnO crystal~\cite{Anderson}, where in the ground state,
unpaired $3d$ electrons on Mn are separated by
O with a filled $2p$ shell. The system takes the form of a three-centre cluster model, as in
Fig.~\ref{superlevels}. The $d-p$ energy difference, $\epsilon_d-\epsilon_p,$ also
known as the charge-transfer energy, has been defined as $\Delta$.

To derive an expression for the exchange interaction
between singly-occupied Mn cation sites, we use a two-band Hamiltonian of the Anderson-Hubbard type, given in
second-quantized form by
\begin{equation}
H=\sum_{i,\sigma}\left(\epsilon_i-\mu\right) n_{i,\sigma}-t\sum_{\langle i,j\rangle,\sigma}
\left(c^\dagger_{i,\sigma}c_{j,\sigma}+c^\dagger_{j,\sigma}c_{i,\sigma}\right)+\sum_i U_i n_{i,\uparrow} n_{i,\downarrow}
\label{H1}
\end{equation}
where $\langle i,j\rangle$ denotes that the hopping with energy $t$ is restricted to nearest neighbours, the $\epsilon_i$ term describes the on-site energy of an electron,
and $n_{i,\sigma}\equiv c^\dagger_{i,\sigma}c_{i,\sigma}$ is the number of electrons of spin $\sigma$ on
site $i$. The presence of two electrons of opposite spin on a $d$ orbital (Mn) comes at an energy cost $U$ due to Coulomb repulsion; $U_i$
for $i$ corresponding to the central site, O, is taken to be zero. Finally, $\mu,$ taken to be equal to $(\epsilon_p+\epsilon_d)/2$,
serves to shift the zero of energy.
The Hamiltonian Eq.~(\ref{H1}) may be depicted in tri-diagonal block form:
\begin{equation}
H=\left(\begin{array}{c|c|c}
H_{00} & T_{01}&0\\
\hline
T_{10} & H_{11}&T_{12}\\
\hline
0&T_{21}&H_{22}
\end{array}\right)
\label{block}
\end{equation}
with the nine anti-ferromagnetic basis configurations
(sufficient for finding exchange) for four electrons on three sites,
\begin{equation}
\begin{array}{c c}
\mathrm{row}\; 0:& |\uparrow,\uparrow\downarrow,\downarrow\rangle\;,|\downarrow,\uparrow\downarrow,\uparrow\rangle\\
\mathrm{row}\; 1:& |\uparrow\downarrow,\uparrow,\downarrow\rangle\;,|\uparrow\downarrow,\downarrow,\uparrow\rangle\;,|\uparrow,\downarrow,\uparrow\downarrow\rangle\;,|\downarrow,\uparrow,\uparrow\downarrow\rangle\\
\mathrm{row}\; 2:& |\uparrow\downarrow,\cdot,\uparrow\downarrow\rangle\;,|\uparrow\downarrow,\uparrow\downarrow,\cdot\rangle\;,|\cdot,\uparrow\downarrow,\uparrow\downarrow\rangle
\end{array}.
\end{equation}
The unperturbed basis functions of block $H_{00}$ have an energy of zero because of shift of the energy zero with $\mu$ as discussed
above.

\subsection{Effective Hamiltonian approach}
\label{sec:EffHam}
We employ the formalism of the effective Hamiltonian, derived \textit{exactly}, for the calculation of superexchange.
The purpose of defining an effective Hamiltonian is to describe the full Hamiltonian
in terms of a small subset of the basis, which for our purposes, is the $H_{00}$ block (see Eq.~(\ref{block})) as
we are interested in
the superexchange interaction between singly-occupied cation orbitals separated by O. An exact projection may be obtained
from a resolvent approach (\textit{e.g.} Ref.~\cite{Lowdin}). The solution, in terms of the notation of Eq.~(\ref{block}),
is given by
\begin{equation}
\underline{\underline{H_{\mathrm{eff}}}}(E)=\underline{\underline{H_{00}}}+\underline{\underline{T_{01}}}\left(E\underline{\underline{I_{11}}}-(\underline{\underline{H_{11}}}+\underline{\underline{T_{12}}}(E\underline{\underline{I_{22}}}-\underline{\underline{H_{22}}})^{-1}\underline{\underline{T_{21}}})\right)^{-1}\underline{\underline{T_{10}}}
\label{Heff}
\end{equation}
where $I_{nn}$ is the identity matrix of the $nn$ diagonal block of Eq.~(\ref{block}) and the matrix notation will be omitted hereafter for clarity.
Our eigenvalue problem thus becomes
\begin{equation}
H_{\mathrm{eff}}(E)X=EX,\;\;\;X =\left|\uparrow\,,\uparrow\downarrow\,,\downarrow\rangle\right.\,\pm\,\left|\downarrow\,,\uparrow\downarrow\,,\uparrow\rangle\right.
\label{eigHeff}
\end{equation}
for the projection onto singlet and triplet combinations of separated $d$ electrons.

The undetermined $E$, appearing on both sides of Eqs.~(\ref{Heff})-(\ref{eigHeff}), can lead to the exact
eigenvalue for \emph{one} element $i$ at a time in $X$ if we solve for the roots of the equation $\lambda_i(E)=E$
for each corresponding eigenvalue $\lambda_i(E)$ of $H_{\mathrm{eff}}(E)$. We essentially
have two $H_{\mathrm{eff}},$ for all of the eigenvalues of the singlet and triplet subspaces respectively.
The difference in the lowest singlet and triplet energies, $E(S)-E(T),$ is defined as the superexchange $J$.

The formalism of the effective Hamiltonian is invoked here for two reasons. The first is
to compare the results with the
frequent implementation of using one effective Hamiltonian to describe all solutions within an
energetically-degenerate subspace~\cite{Lindgren,BWbook} by approximating the unknown energy $E$ in Eqs.~(\ref{Heff})-(\ref{eigHeff})
by the unperturbed energy, as in Rayleigh-Schr\"{o}dinger perturbation theory. When
 followed by an expansion for small $t$, this leads to the same result as a $T$-matrix expansion with
the unperturbed energy. The limitations
of using unperturbed energies in Brillouin-Wigner perturbation theory to derive
effective Hamiltonians have been previously noted~\cite{GangSu}. The correspondence
between resolvent methods, which we have used, and Brillouin-Wigner perturbation theory,
in the derivation of effective Hamiltonians
has been discussed extensively in \textit{e.g.} Refs.~\cite{Lowdin} and~\cite{Chernyshev}. The second reason for projecting onto
the model subspace is technical: the particular splitting beforehand into two symmetry subspaces, singlet
and triplet, enabled
\textsc{Mathematica}~\cite{math} to find the required roots in closed form.
\section{Results}
\label{sec:results}
 In this section, the exact, closed-form expressions for $E(S),$ $E(T),$ and $J$ provided by
the effective-Hamiltonian formalism will be expanded in various limiting forms.
\subsection{Limiting forms for $U\gg t$}
The expansion of the exact lowest triplet and singlet energy eigenvalues for $U\gg t$ up to $O(t^4)$ is
\begin{eqnarray}
E(T)&=&\left(\frac{2\Delta^3}{U^4}-\frac{2\Delta^2}{U^3}+\frac{2\Delta}{U^2}-\frac{2}{U}\right)t^2
+\left(\frac{4}{U^3}-\frac{12\Delta}{U^4}\right)t^4+O(t^6)\;\;\;\nonumber\\
E(S)&=&\left(\frac{2\Delta^3}{U^4}-\frac{2\Delta^2}{U^3}+\frac{2\Delta}{U^2}-\frac{2}{U}\right)t^2
+\left(\frac{8\Delta}{U^4}-\frac{4}{U^3}\right)t^4+O(t^6)
\label{Uggt}
\end{eqnarray}
such that
\begin{equation}
J=\frac{4t^4(5\Delta-2U)}{U^4}.
\label{JUt}
\end{equation}
Equations~(\ref{Uggt})-(\ref{JUt}) were derived by substituting for $U\rightarrow t/x$ in the exact $J$ and
expanding about $x.$ The
same result is obtained for $U\rightarrow\infty$ for up to $O(1/U^4)$ (single-scale limit).

On the other hand, setting $t\rightarrow Ux$ gives a
totally different result:
\begin{equation}
J=\frac{-4(\Delta+2U)t^4}{U(\Delta+U)^3}
\label{goodJ}.
\end{equation}
The expression~(\ref{goodJ}) agrees with that obtained from expanding the exact $J$ in
the single-scale series in $t\rightarrow 0,$ but only up to fourth order - at higher order the expressions differ. The
result~(\ref{goodJ}) is widely used for the $t\ll U$ situation (see \textit{e.g.} Eq. 13 in Ref.~\cite{Jefferson1}, and
Eq. 16 in Ref.~\cite{Manousakis}), and it is often used as is, to describe either the M-H case
$\Delta>U,$~\cite{Zaanen1,Jefferson1,Jefferson2} or the C-T
case $\Delta<U$~\cite{Zaanen1,Manousakis} without further, unrealistic limits relating $\Delta$ to $U$ considered.

\begin{figure}[htb]
\includegraphics[width=7.cm]{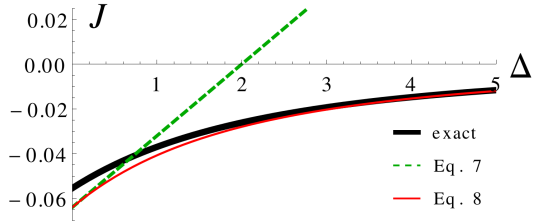}
\caption{For fixed $t=1, U=4,$ the variation of $J$ with $\Delta$ in the exact form, and
$t\ll U$ according to the two approximation methods, leading to Eqs.~(\ref{JUt}) and~(\ref{goodJ}) respectively.}
\label{Eq7-8}
\end{figure}

In deriving Eqs.~(\ref{JUt})-(\ref{goodJ}),
no assumptions have been made for $\Delta$ and yet, as can be seen from Fig.~\ref{Eq7-8},
although the two expressions coincide for very small $\Delta,$
for increasing $\Delta$ Eq.~(\ref{JUt}) poorly represents $J$ as it increases without bound (except for the trivial
case $U=\infty$ where it correctly gives zero). On the other hand, the convergence of the
expansion leading to Eq.~(\ref{goodJ}) improves as $\Delta$ is increased. Thus, even though Eq.~(\ref{goodJ})
seems better than Eq.~(\ref{JUt}), in both cases, the convergence of the $t/U$ expansion
is impacted by the value of $\Delta.$ This concept of parameter-dependent convergence and its consequences
will be discussed in detail in Sec.~\ref{sec:ROC}.

If, instead of using the exact $E$ in Eq.~(\ref{Heff}), the
unperturbed energy is used, as described in Sec.~\ref{sec:EffHam}, we obtain, without further approximations,
\begin{equation}
J=\frac{-4 t^4 (\Delta+2 U)}{(\Delta+U) \left(\Delta^2 U+2 \Delta \left(U^2-t^2\right)-4 t^2 U+U^3\right)}
\label{JUt2}.
\end{equation}
The same expressions as Eq.~(\ref{JUt}) or~(\ref{goodJ}) are obtained - with the agreement stopping after
fourth order - upon setting $U\rightarrow t/x$ or $t\rightarrow Ux$
respectively and expanding about $x=t/U$ (or by defining $x=t/\Delta$). When $t\rightarrow Ux$ is set, the results are exactly
the same to all orders as a T-matrix expansion with the unperturbed energy. On the other hand,
 taking the series expansion $U\ll\Delta$ of Eq.~(\ref{JUt2}) does not
lead to the results presented in Sec.~\ref{sec:dggu} for the $U\ll\Delta$
limit of the exact $J$; the validity of the perturbation expansion
requires that $t$ be ``much smaller" than all of the other energy scales
in the problem and the convergence of $U\ll\Delta$ is mostly incompatible with this.

\subsection{Anderson expression}
\label{sec:anderson}
The usual ``Anderson" expression for superexchange~\cite{Anderson} can be
derived by considering the $U\ll\Delta$ limit of expression~(\ref{goodJ}),
yielding, in terms of the direct-exchange term $J_{dir}=-4t^2/U$ connecting $d-d$
nearest neighbours:
\begin{equation}
J_\mathrm{Anderson}=\frac{J_{dir}t^2}{\Delta^2}.
\label{anderson}
\end{equation}
Equation~(\ref{anderson}) was obtained with $\Delta\rightarrow U/x$
and expanding about $x$ up to second order. On the other hand, the substitution $U\rightarrow \Delta x$
and expanding about $x$ up to second order. On the other hand, the substitution $U\rightarrow \Delta x$
into Eq.~(\ref{goodJ}) yields the modified expression, to first order,
\begin{equation}
J_\mathrm{Anderson\; 2}=\frac{J_{dir}t^2(\Delta-U)}{\Delta^3}
\label{anderson2}
\end{equation}
(see also Ref.~\cite{feldkemper} Eq.~7 in slightly different notation) which leads Eq.~(\ref{anderson}) by two orders of expansion. The result~(\ref{anderson2})
also arises from setting $t\rightarrow xU$
followed by $U\rightarrow y\Delta$ in the exact $J$, and expanding in $x$ and $y$ simultaneously,
indicating that the neglect of mixed derivatives does not affect the result. However, the result remains intrinsically
biased by the form of the substitution: expanding the series separately entails substituting back for the first
expansion variable before proceeding to the second expansion.

\subsection{$U\ll \Delta$ expansion}
\label{sec:dggu}
\begin{figure}[htb]
\includegraphics[width=7.cm]{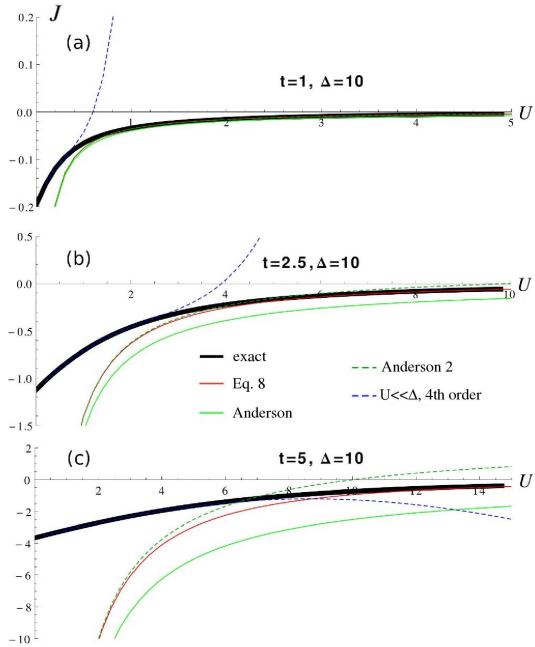}
\caption{The variation of $J$ with $U$ for fixed $\Delta=10$ and different values of $t$, in the exact form and in the
limiting cases $t\ll U$ (Eq.~(\ref{goodJ})), the $U\ll\Delta$ limit of $t\ll U$ (``Anderson" case, (Eq.~(\ref{anderson})))
and ``Anderson 2" (Eq.~(\ref{anderson2})), and $U\ll\Delta$ as discussed in the text.}
\label{dggu}
\end{figure}

Rather than begin with small $t,$ we first expand the exact $J$ in $U\ll\Delta$ using $U\rightarrow \Delta x$
as in the ``Anderson 2" expression Eq.~(\ref{anderson2}). We expand up to $O(x^4)$. The result, although
in closed form, is very unwieldy. For large $t$ the expansion is accurate,
but for small $t$ it diverges for all but the smallest
values of $U$ as can be seen from Fig.~\ref{dggu}. This shows
explicitly that the $U/\Delta$ expansion is parameter-dependent. The
$t\ll U$ expansion (Eq.~(\ref{goodJ})) is also shown for comparison, even though it is not expected
to hold for $t>U$. Extrapolating beyond (c) for even larger $t,$ the $U/\Delta$ expansion
is found to be very accurate, even for $U>\Delta$.

The limiting forms Eqs.~(\ref{goodJ}),~(\ref{anderson})-(\ref{anderson2}), provided that
$t\ll U\ll\Delta$ is strictly adhered to, lead to an accurate result for $J$. However, the meaning of the term
`strictly' here is vague because, as will be discussed in greater detail in Sec.~\ref{sec:ROC},
the two series expansions are correlated and this impacts the convergence, causing it
to be parameter-dependent.

\begin{figure}[htb]
\includegraphics[width=7.cm]{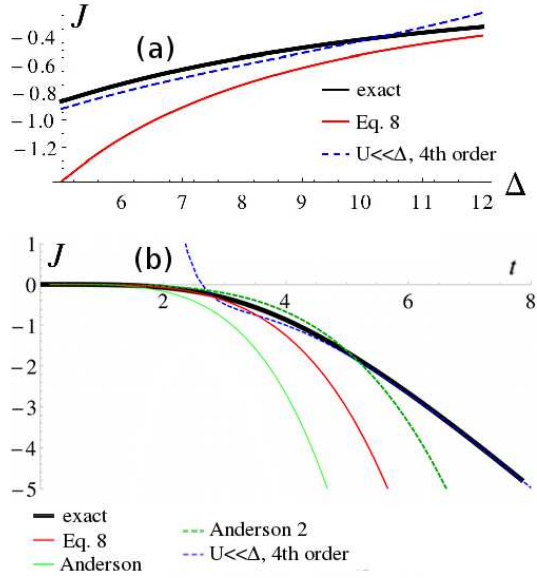}
\caption{(a) the variation of $J$ with $\Delta$ for $t=3$, $U=4$, and (b) the variation of $J$ with $t$ for $U=6,$ $\Delta=8.$
The various curves denote the exact form, the limiting cases for $t\ll U$ (Eq.~(\ref{goodJ})),
the Anderson expressions Eqs.~(\ref{anderson}) and~(\ref{anderson2}), and $U\ll \Delta.$}
\label{Jd}
\end{figure}

These limiting forms are worse than the $U\ll\Delta$ expression for cases in which the
milder relationship $t\lesssim U\lesssim\Delta$ holds. Here, the $U\ll\Delta$ expansion coincides
with the exact value. For
instance in Fig.~\ref{dggu}(c), a value of $J=-3.5$ is reached with $U$=7.1, 4.5, and 0.3 for the Anderson, Eq.~(\ref{goodJ}),
and $U\ll\Delta$ expressions respectively, causing the M-H relationship $t<U<\Delta$ to be fictitiously
satisfied with the use of the Anderson expression. $U$ obtained from the exact $J$ is the same
as that from the fourth-order $U\ll\Delta$ expression for this case. Similarly, for Fig.~\ref{dggu}(b), for a measured $J=-0.3$, the values
are $U$=5.2, 3.6, and 3.1 respectively, such that the qualitative
relationship $t<U<\Delta$ is correctly determined, but with an inaccurate $U$ in the Anderson model. Fig.~\ref{Jd}(a)
shows another perspective, that for $t\lesssim U$ the fitting of $J$ to the model of Eq.~(\ref{goodJ}), as
opposed to the $U\ll\Delta$ limiting form, can give a
value for $\Delta$ much larger than the correct one. Yet another depiction (Fig.~\ref{Jd}(b)) shows how the deduced $t$ can
be far from the correct value when $t\lesssim U\lesssim \Delta,$ and more importantly, that expanding the exact $J$ with
$U\ll\Delta$ is more accurate than using $t\ll U.$

\begin{figure}[htb]
\includegraphics[width=10.cm]{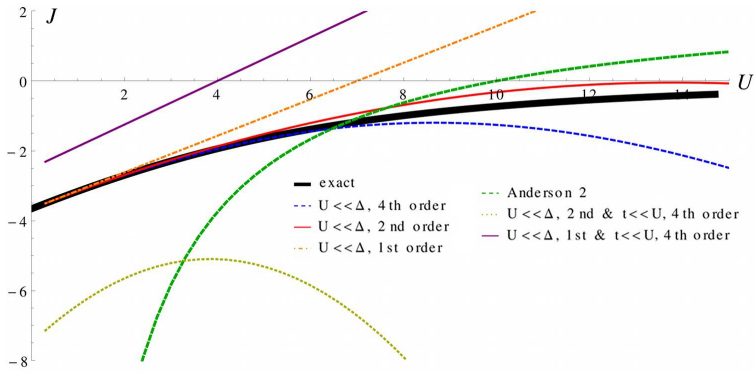}
\caption{The variation of $J$ with $U$ for $t=5$ and $\Delta=10,$ (as in Fig.~\ref{dggu}(c)) in the exact form and in various
limiting cases. Specifically, expansions beginning with the $U\ll\Delta$ limit, some followed by a further expansion
with $t\ll U,$ are shown alongside the ``Anderson 2" expression, Eq.~(\ref{anderson2}), which arises from the reverse order of
two expansions: $t\ll U$ first and then $U\ll\Delta.$ See text for more details.}
\label{dUcompare}
\end{figure}

If the limits $U\ll\Delta$
and $t\ll U$ were in fact interchangeable, then Eq.~(\ref{anderson2}), which was also derived with $U\rightarrow \Delta x,$
would agree with the corresponding curve (first-order expansion in $U/\Delta$ followed by fourth-order
expansion in $t/U$) of Fig.~\ref{dUcompare}. In both cases, the variable substitutions  were performed in exactly the same
way - only the order of the expansions is reversed. It is clear from Fig.~\ref{dUcompare} that the convergence with increasing
order cannot be ensured, something found to occur also with other parameter values and types of
depictions. The second and fourth-order $U/\Delta$ expansions, without further limits taken, are quite reliable for certain regions of
parameter space, often better than the $t/U$ expansions, but for lack of representation by simple expressions they have to be used in their
cumbersome forms. Although the $t/U$ expansions of the $U\ll\Delta$ curves do have simple forms, such expressions would
be worthless, as can be seen
from Fig.~\ref{dUcompare}. Incidentally, neither in the C-T insulator $t\ll\Delta\ll U$ situation, do the limits commute and similar
conclusions can be made.

\begin{figure}[htb]
\includegraphics[width=7.5cm]{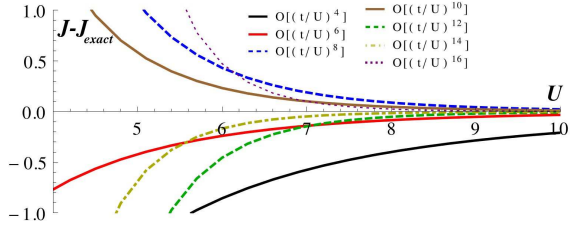}
\caption{The difference of the approximate $J$ in its various orders of expansion in $t/U$ (or $t/\Delta,$ with the same result)
from the exact value, as a function of $U$ for fixed $t=5$, $\Delta=10$ (as in Figs.~\ref{dggu}(c) and~\ref{dUcompare}). }
\label{diverg0}
\end{figure}

The disagreement in the results depending on the sequence in which the expansions are taken
means that the limit is formally undefined. This manifests itself
as a better agreement of the $U\ll\Delta$ expression with the exact result for the milder limits compared with
the results derived starting with $t\ll U,$ due to a divergence of the latter with increasing expansion order
in this region (Fig.~\ref{diverg0}).

\subsection{Radius of convergence}
\label{sec:ROC}
The inconsistency between the two types of expansion
procedures (\textit{e.g.} leading to Eqs.~(\ref{JUt}) and~(\ref{goodJ}) respectively) is
due to the radius of convergence (ROC) being unstable due to the
multiple variables and due to the fractional powers present in the function. Consider the
triplet energy, which has the simple form
\begin{equation}
E(T)=\frac{1}{2} \left(\Delta+U-\sqrt{\Delta^2+2 \Delta U+8 t^2+U^2}\right).
\end{equation}
Substituting for $t$ as $t\rightarrow xU$ yields
one form for the series, while another form is found with $U\rightarrow t/x$. The first form
has a ROC which is dependent on $\frac{(U+\Delta)^2}{U^2},$
while in the second form there is an unstable parameter dependence with increasing
series order. The ROC of both forms coincide and are parameter-free when
$\Delta=0$. The parameter-dependence of the ROC
is the root cause of the limits $t\ll U$ and $U\ll\Delta$ of $J$ being non-commuting; even
in the simultaneous limits described earlier at the end of Sec.~\ref{sec:anderson},
the result was biased by the
form of the substitution. The limits are ``correlated" and this impacts
the convergence condition.
\begin{figure}[htb]
\includegraphics[width=7.5cm]{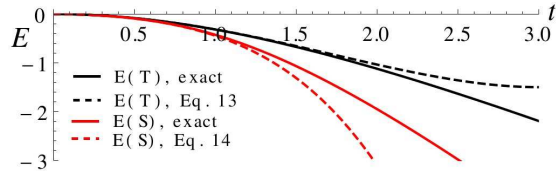}
\caption{For fixed $U=1$, $\Delta=5$, the variation of the triplet
and singlet eigenvalues with $t$ in the exact form and in the
approximate forms Eqs.~(\ref{ET})-(\ref{ES}).}
\label{conv}
\end{figure}

Parameter-dependent ROC require that
an additional condition be satisfied, in addition to the smallness of the expansion parameter.
Together they yield an expansion variable $x^*=t/(\Delta+U)$ for $E(T)$ which yields
global (parameter-free) convergence
for $x^*< \frac{1}{2\sqrt{2}}$. In the case of the singlet however,
the ROC remains parameter-dependent and more restrictive. Up
to fourth order in the expansion parameter $x^*$, we have deduced the expressions:
\begin{eqnarray}
E(T)&=&-\frac{2t^2}{U+\Delta}+\frac{4t^4}{(U+\Delta)^3}\label{ET}\\
E(S)&=&-\frac{2t^2}{U+\Delta}-\frac{4t^4}{U(U+\Delta)^2}\label{ES},
\end{eqnarray}
which result again in Eq.~(\ref{goodJ}) for $J$, a ubiquitous expression also obtained with various other
expansion parameters, such as $x=t^2/(U\Delta).$ It should be noted that the agreement between
the various $J$ expressions obtained with different expansion parameters is fortuitous, occurring
only up to fourth-order in the expansion parameter; at higher order they differ. For $U=1$,
$\Delta=10,$ $x^*$ gave $t/(U+\Delta)\lesssim$ 0.18,
or $t<2.0$ for the singlet, compared with $t<3.9$ for the triplet. Thus, $U$ does not need
to be larger than $t$
for Eq.~(\ref{goodJ}) to be accurate (see
also Fig.~\ref{dggu}(a) for this case). Figure~\ref{conv} shows
the difference in the convergence for $U=1$ and $\Delta=5$ for which the numerical
evaluation gave $t<2.1$ and $t<1.3$ for the convergence condition of the infinite series
expanded in $x^*$ for $E(T)$ and $E(S)$ respectively. Clearly,
$J$ is composed of two functions with different convergence properties.
\section{Conclusions}
\label{sec:conclusions}
We have shown that the commonly referred-to Mott-Hubbard (M-H)
insulator situation, $t\ll U\ll \Delta$ yields the Anderson form
for superexchange only if the limits are taken in a particular sequence. Hence, the limit is actually
undefined (also true for the charge-transfer (C-T) insulator) and the reason
for this is the incompatibility of the expansion variables with the radius of convergence,
which depends on the parameters appearing in the problem. The Anderson expression, or other
limiting expressions, derived \textit{e.g.} from $t\ll U,$
but also from many other parameter relationships,
 agree very well with the exact
result in the extreme form of these limits. However, for milder relationships between
the parameters, \textit{e.g.} $t<U<\Delta,$
the limiting form $t\ll U$ can be significantly less accurate than the oppositely-derived
limit $U\ll\Delta.$

The M-H limits, which arise
from the convenience of the application of perturbation theory,
represent a subset of the range of applicability of the limiting form as
we found that other expansions, which assume different
relationships between the variables, \textit{e.g.} $U<t,$ yield the same limiting
form of $J$ to great success.
We additionally show that since the triplet and singlet energies comprising $J$
display different convergence properties, a single expansion
variable for determining $J$ does not formally exist.

The wider implications of our findings are that the results determined for electronic
parameters, and the quantities derived from them such as electron
transfer rates,
could be questionable if an approximate expression for $J$ instead of the exact result is relied upon. This
is because the
parameters found from fitting to experimental data
could be outside the radius of convergence of the series, particularly
when the parameters do not hold extreme relationships amongst themselves and this may
yield a fictitious qualitative relationship between the parameters. As in previous works,~\cite{LeeOh,Bocquet92,Bocquet}
where more detailed calculations, taking into account higher-energy excitations, have been shown to modify the
characterization (\textit{e.g.} from M-H to C-T insulators) of some transition-metal oxides, here as well,
through exact solutions of a simpler model system, we are able to show that the classification of materials obtained
by models based on perturbative approaches can be modified in regimes where such methods break down.

\section{Acknowledgments}
The author thanks the Crete Center for Theoretical Physics (CCTP) for providing computational
facilities, Dr. G. Kastrinakis for critically reading the
manuscript, and Prof. I. K. Kominis for useful discussions. 
research was partially supported
by the European Union's Seventh Framework
Programme (FP7-REGPOT-2012-2013-1) under grant agreement n$^\mathrm{o}$ 316165.

\end{document}